\documentclass[%
 aip,
 amsmath,amssymb,
 reprint,%
]{revtex4-1}

\usepackage{graphicx}% Include figure files
\usepackage{dcolumn}% Align table columns on decimal point
\usepackage{bm}% bold math
\usepackage[utf8]{inputenc}
\usepackage[T1]{fontenc}
\usepackage{mathptmx}
\usepackage{etoolbox}

\makeatletter
\def\@email#1#2{%
 \endgroup
 \patchcmd{\titleblock@produce}
  {\frontmatter@RRAPformat}
  {\frontmatter@RRAPformat{\produce@RRAP{*#1\href{mailto:#2}{#2}}}\frontmatter@RRAPformat}
  {}{}
}%
\makeatother
\begin{document}

\preprint{AIP/123-QED}

\title{3+1 formulation of light modes in nonlinear electrodynamics}
% Force line breaks with \\
\author{Chul Min Kim}
\email{chulmin@gist.ac.kr}
\affiliation{Advanced Photonics Research Institute, Gwangju Institute of Science and Technology, Gwangju 61005, Korea}
\affiliation{Center for Relativistic Laser Science, Institute for Basic Science, Gwangju 61005, Korea}
\affiliation{Department of Physics and Photon Science, Gwangju Institute of Science and Technology, Gwangju 61005, Korea}

\author{Sang Pyo Kim}
\email{sangkim@kunsan.ac.kr}
\affiliation{Department of Physics, Kunsan National University, Gunsan 54150, Korea}
\affiliation{Asia Pacific Center for Theoretical Physics, Pohang 37673, Korea}

\date{\today}% It is always \today, today,

\begin{abstract}
We present a 3+1 formulation of the light modes in nonlinear electrodynamics described by Plebanski-type Lagrangians, which include Post-Maxwellian, Born-Infeld, ModMax, and Heisenberg-Euler-Schwinger QED Lagrangians. In nonlinear electrodynamics, strong electromagnetic fields modify the vacuum to acquire optical properties. Such a field-modified vacuum can possess electric permittivity, magnetic permeability, and magneto-electric response, inducing novel phenomena such as vacuum birefringence. By exploiting the mathematical structures of Plebanski-type Lagrangians, we obtain a streamlined procedure and explicit formulas to determine light modes, i.e., refractive indices and polarization vectors for a given propagation direction. We also work out the light modes of the mentioned Lagrangians for an arbitrarily strong magnetic field. The 3+1 formulation advanced in this paper has direct applications to the current vacuum birefringence research: terrestrial experiments using permanent magnets/ultra-intense lasers for the subcritical regime and astrophysical observation of the x-rays from highly magnetized neutron stars for the near-critical and supercritical regimes.
\end{abstract}

\maketitle

%\begin{quotation}
%The ``lead paragraph'' is encapsulated with the \LaTeX\ 
%\verb+quotation+ environment and is formatted as a single paragraph before the first section %heading. 
%(The \verb+quotation+ environment reverts to its usual meaning after the first sectioning command.) 
%Note that numbered references are allowed in the lead paragraph.
%
%The lead paragraph will only be found in an article being prepared for the journal \textit{Chaos}.
%\end{quotation}

\section{Introduction} \label{sec:intro}

Light propagation has been an interesting theoretical issue in curved spacetimes.\cite{jordan2013republication}
Recently the Event Horizon Telescope (EHT) observed the shadow of supermassive black holes from light propagating around the event horizon and provided a way to understand the geometry of black holes. \cite{Akiyama2019first,Akiyama2024first,Vagnozzi2023Horizon} Nonlinear electrodynamics has also provided nontrivial background for light propagation. Born and Infeld introduced a nonlinear Lagrangian of the Maxwell scalar and its pseudo-scalar, which recovers the Maxwell theory in the weak field limit but significantly modifies the physics of light-matter and light-light interactions in the strong field limit.\cite{Born1934Foundations} The causality of Born-Infeld Lagrangian was studied.\cite{gibbons2001born} Heisenberg and Euler obtained the exact one-loop effective Lagrangian for electrons in a constant electromagnetic field.\cite{Heisenberg1936Folgerungen} Later, Schwinger introduced the proper-time method for quantum field theory and obtained the one-loop effective Lagrangian for spinless charged bosons and spin-1/2 fermions in a constant electromagnetic field.\cite{Schwinger1951Gauge}

The prominent features of Heisenberg-Euler-Schwinger (HES) QED Lagrangian are the vacuum polarization \cite{Dittrich1985Effective,Fradkin1991Quantum} and the so-called (Sauter-)Schwinger pair production
of electrons and positrons.\cite{Sauter1931Uber,Schwinger1951Gauge} Interestingly, the vacuum becomes unstable in strong electric fields due to pair production, which is a consequence of the imaginary part of the HES Lagrangian.\cite{Kim2008Effective,Kim2010Effective} The polarized vacua in strong electromagnetic fields provide a probe photon (light) with nontrivial backgrounds. In recent years, nonlinear electrodynamics (NED) has been intensively studied partly because ultra-intense lasers using the chirped pulse amplification (CPA) technology have been developed \cite{Strickland1985Compression,Danson2019Petawatt} and partly because highly magnetized neutron stars and magnetars have been observed; these astrophysical bodies have magnetic fields comparable to or stronger than the critical field strength $B_c = m^2 c^2/e \hbar = 4.4 \times 10^{13}\, {\rm G}$.\cite{Kaspi2017Magnetars} An electromagnetic field makes the vacuum a polarized medium, and the polarized vacuum behaves similarly as the dielectric or ferromagnetic medium. Nonlinear electrodynamics in general exhibits, in addition to electric permittivity and magnetic permeability, the magneto-electric effect, in which a magnetic field induces electric polarization whereas an electric field induces magnetization.\cite{spaldin2005renaissance,Eerenstein2006Multiferroic,spaldin2019advances} A strong electric field, on the other hand, creates pairs of charged particles and anti-particles, known as the (Sauter-)Schwinger pair production.

Nonlinear electrodynamics thus exhibits a rich structure of vacuum polarization, such as vacuum birefringence, photon propagation, and magneto-electric effect. In a strong magnetic field, the polarized vacuum acquires non-trivial refractive indices, which causes birefringence in the photon propagation either along the magnetic field or in the perpendicular plane.\cite{BialynickaBirula1970Nonlinear,Adler1971Photon} Light propagation has been intensively studied in NED of Plebanski Lagrangians,\cite{plebanski1970lectures} in which a probe photon of low energy obeys a wave equation in nonlinear backgrounds. Most literature has employed the covariant four-vector formulation of the light propagation (for instance, see references\cite{BialynickaBirula1970Nonlinear,Dittrich1998Light,Lorenci2000Light,Karbstein2015Photon,Obukhov2002Fresnel,Rubilar2002Linear}), while the 3+1 formulation of photon propagation was used for the post-Maxwellian theory in weak electromagnetic fields.\cite{Ni2013Foundations}

In a previous study, we investigated the light propagation in a supercritical magnetic field and a weak electric field, especially in the case of a non-zero electric field along the magnetic field (electromagnetic wrench).\cite{Kim2021Magnetars,Kim2023Vacuum} In contrast to the pure magnetic field case, the electromagnetic wrench introduces magneto-electric effects and thus significantly modifies the light modes. In our previous reports,\cite{Kim2021Magnetars,Kim2023Vacuum} we focused on the configuration of parallel electric and magnetic fields to study the effect in the simplest setting. However, considering complicated field configurations around neutron stars or focal regions of ultra-intense lasers, we need to extend the previous formulation to an arbitrary field configuration. Furthermore, it is worthy to make the formulation applicable to other Lagrangians considered in the context of general NED.

In this paper, we introduce the 3+1 formulation of light modes in the NED of the Plebanski-type Lagrangians ${\cal L} (F, G)$,\cite{plebanski1970lectures} where $F$ is the Maxwell scalar, and $G$ is the Maxwell pseudo-scalar.
\begin{equation} \label{FG}
	F\equiv\frac{1}{4}F^{\mu\nu}F_{\mu\nu}=\frac{1}{2}\left(\mathbf{B}^{2}-\mathbf{E}^{2}\right),\ G\equiv\frac{1}{4}F^{\mu\nu}F_{\mu\nu}^{*}=-\mathbf{E}\cdot\mathbf{B}.
\end{equation}
Here, $F^{\mu\nu} = \partial^{\mu} A^{\nu} - \partial^{\nu} A^{\mu}$ is the field-strength tensor, and $F^{*\mu\nu}=\varepsilon^{\mu\nu\alpha\beta}F_{\alpha\beta}/2$ ($\varepsilon^{0123}=1$) is the dual field-strength tensor.\cite{Jackson1999Classical} Also provided is a streamlined procedure for determining light modes with explicit formulas, which should be valuable to laboratory experiments and astrophysical observations of vacuum birefringence. As concrete examples, we apply the formulation to the frequently used NED Lagrangians for an arbitrarily strong magnetic field to present the expressions of light modes. Those expressions can be used as optical tools to test theoretical black holes with NED Lagrangians. We use the Lorentz-Heaviside units with $c=\hbar=1$, and thus $\alpha_e=e^{2}/4\pi$, where $e$ is the elementary charge.

In the Minkowski spacetime with the metric $(+,-,-,-)$, the timelike Killing vector $\partial_t$ makes the spacetime manifold foliate into one-parameter spacelike hypersurfaces. On each hypersurface $\Sigma_t$, the energy-momentum of a photon is given by $k^{\mu} = (\omega, {\bf k})$  and the field by $F^{\mu\nu}$. In curved spacetime, the 3+1 formulation of Maxwell theory\cite{Komissarov2004Electrodynamics,Kandus2011Primordial,Gralla2014Spacetime} has often been preferred for modeling astrophysics and cosmology. In the 3+1 formulation, the propagation and polarization of probe photons (light) are expressed in terms of directly measurable electromagnetic fields $(\mathbf{E},\mathbf{B})$ and the direction of photon $\mathbf{k}$ on each $\Sigma_t$. In practice, the 3+1 formulation has the advantage of providing the light modes in general NED in the familiar language of conventional optics, facilitating the design of experiments and observations.

This paper is organized as follows. In Section \ref{sec:framework}, we present the general 3+1 framework for analyzing the light modes in the media under strong background fields. Then, in Section \ref{sec:me_material_science}, it is compared with the formulation used in material science to deal with the magneto-electric effect. In Section \ref{sec:3+1}, we apply the framework to the NED of Plebanski-type Lagrangians and compare it with the alternative covariant formulation. In Section \ref{sec:models}, we present specific examples with the representative NED Lagrangians: the Post-Maxwellian (PM), Born-Infeld (BI), ModMax (modified Maxwell, MM), and Heisenberg-Euler-Schwinger (HES) QED Lagrangians. For these Lagrangians, the expressions of the light modes under an arbitrarily strong magnetic field are worked out.  Finally, in Section \ref{sec:conclusion}, summarizing the results, we make a conclusion, discussing the implications of the 3+1 formulation to the vacuum birefringence research, e.g.,  the terrestrial experiments using ultra-intense lasers in the weak field regime and the astrophysical observations for highly magnetized neutron stars in the strong field regime. 

%TO BE USED ELSEWHERE
%On each $\Sigma_t$ whose observers measure the electromagnetic fields $\{ {\bf E}, {\bf B} \}$ and the photon's propagation 4-vector $(\omega, {\bf k})$, the constitutive equations $\{ {\bf D} ({\bf E, B}), {\bf H} ({\bf E, B}) \}$  are derived. According to the formulation, the probe photon (light) experiences the effects of not only electric permittivity and magnetic permeability but also magneto-electric responses. Then, expressing the constitutive equation for the probe photon in terms of dyadics, we obtain the complete information about the light modes, i.e., refractive indices and polarization vectors. 
%Also as found in Secsion \ref{sec:comp_cov}, we compare the results from the 3+1 formulation with those from the covariant formulation in the literature. These two formulations agree with each other, while the 3+1 formulation is more convenient for observations or measurements.

\section{3+1 framework of analyzing the light propagation modes in a media under strong background fields} \label{sec:framework}
In this section, we set up a 3+1 framework to obtain the light propagation modes in arbitrary media under strong background fields. When applied to NED Plebanski-type Lagrangians, as shown in Section \ref{sec:3+1}, it yields refractive indices and polarization vectors for a given propagation direction. 

In classical electrodynamics, the electromagnetic properties of media are described in terms of the polarization $\boldsymbol{\mathcal{P}}$, the magnetization  $\boldsymbol{\mathcal{M}}$  the electric displacement $\mathbf{D}$, and $\mathbf{H}$:\cite{Landau1984Electrodynamics} 
\begin{equation} \label{DH}
	\mathbf{D}=\mathbf{E}+\boldsymbol{\mathcal{P}},
	\quad
	\mathbf{H}=\mathbf{B}-\boldsymbol{\mathcal{M}}.
\end{equation}
To find the modes of weak low-frequency\footnote{The probe light's frequency $\omega$ is assumed to be much smaller than the medium's characteristic frequency. For the QED vacuum, the characteristic frequency is associated with the electron's rest mass energy $m \simeq 0.5\ \mathrm{MeV}$.}  probe light  ($\delta\mathbf{E}$, $\delta\mathbf{B}$) in the medium polarized and magnetized by a strong slowly-varying background electromagnetic field  ($\mathbf{E}_{0}$, $\mathbf{B}_{0}$), we decompose $\mathbf{E}$, $\mathbf{B}$, $\mathbf{D}$, $\mathbf{H}$, $\boldsymbol{\mathcal{P}}$, and $\boldsymbol{\mathcal{M}}$ as
\begin{equation}
	\mathbf{A}=\mathbf{A}_{0}+\delta\mathbf{A},
\end{equation}
where $\mathbf{A}_{0}$ ($\delta\mathbf{A}$) refers to the background (probe) part.\cite{Mckenna1963Nonlinear,Adler1971Photon} The background part $\mathbf{A}_{0}$ is assumed uniform (locally constant) over the characteristic scales of the probe part  $\delta\mathbf{A}$, which allows us to argue that the relations (\ref{DH}) should be satisfied separately for the background and  probe parts:
\begin{equation} \label{decomposition}
	\begin{split} \mathbf{D}_{0}=\mathbf{E}_{0}+ \boldsymbol{\mathcal{P}}_{0},\quad\mathbf{H}_{0}=\mathbf{B}_{0}- \boldsymbol{\mathcal{M}}_{0}, \\
	\delta\mathbf{D}=\delta\mathbf{E}+\delta \boldsymbol{\mathcal{P}},\quad\delta\mathbf{H}=\delta\mathbf{B}-\delta\boldsymbol{\mathcal{M}}.
	\end{split}
\end{equation}

Because the probe light is weak, only the linear response to the probe light is considered:
\begin{equation} \label{dDdH_dEdB}
	\delta\mathbf{D}=\boldsymbol{\epsilon}_{E}\cdot\delta\mathbf{E}+\boldsymbol{\epsilon}_{B}\cdot\delta\mathbf{B},\quad
	\delta\mathbf{H}=\bar{\boldsymbol{\mu}}_{B}\cdot\delta\mathbf{B}+\bar{\boldsymbol{\mu}}_{E}\cdot\delta\mathbf{E},
\end{equation}
where $\boldsymbol{\epsilon}_{E}$, $\boldsymbol{\epsilon}_{B}$, $\bar{\boldsymbol{\mu}}_{B}$, and $\bar{\boldsymbol{\mu}}_{E}$ (rank-2 tensors in the 3D space) are the permittivity ($\boldsymbol{\epsilon}_{E}$), permeability ($\bar{\boldsymbol{\mu}}_{B}$), and magneto-electric response ($\boldsymbol{\epsilon}_{B}$ and  $\bar{\boldsymbol{\mu}}_{E}$) tensors. These tensors, evaluated with the condition  $(\mathbf{E},\mathbf{B})=(\mathbf{E}_0,\mathbf{B}_0)$, represent the influence of the strong background fields on the probe light.  The tensors $\boldsymbol{\epsilon}_{B}$ and $\bar{\boldsymbol{\mu}}_{E}$, absent in the theory of usual optical materials,\cite{Born1999Principles} represent the magneto-electric effect: the magnetic field induces polarization, and the electric field induces magnetization.\cite{Melrose1991Electromagnetic} 

When there is no real charge and current, the Maxwell equations for the probe light become 
\begin{equation} \label{Maxwell}
	-\frac{\partial\delta\mathbf{B}}{\partial t}=\nabla\times\delta\mathbf{E}, \quad \frac{\partial\delta\mathbf{D}}{\partial t}=\nabla\times\delta\mathbf{H},
\end{equation}
which reduce to
\begin{equation} \label{plane wave}
	\delta\mathbf{B}=\mathbf{n}\times\delta\mathbf{E},\quad
	\delta\mathbf{D}=-\mathbf{n}\times\delta\mathbf{H}
\end{equation}
for a Fourier mode characterized by an angular frequency $\omega$ and a propagation vector  $\mathbf{k}=\omega\mathbf{n}=\omega n \hat{\mathbf{n}}$;  $n$ is the refractive index of the mode, and  $\delta\mathbf{B}$,  $\delta\mathbf{E}$,  $\delta\mathbf{D}$, and  $\delta\mathbf{H}$  vary as $\exp(i\omega(\mathbf{n}\cdot\mathbf{x}-t))$.

Then, substituting (\ref{plane wave}) into (\ref{dDdH_dEdB}) leads to the master equation that determines the refractive indices and polarization vectors ($n$ and $\delta\mathbf{E}$) for a given propagation direction ($\hat{\mathbf{n}}$):
\begin{equation} \label{Lambda}
\boldsymbol{\epsilon}_{E}\cdot\delta\mathbf{E}+\boldsymbol{\epsilon}_{B}\cdot\mathbf{n}\times\delta\mathbf{E}
+\mathbf{n}\times\left(\bar{\boldsymbol{\mu}}_{B}\cdot\mathbf{n}\times\delta\mathbf{E}+\bar{\boldsymbol{\mu}}_{E}\cdot\delta\mathbf{E}\right)=0.
\end{equation}
The equation can be written in a matrix-vector form:
\begin{equation} \label{Lambda_eps_mu}
	\boldsymbol{\Lambda}\cdot \delta\mathbf{E}=\left( \boldsymbol{\epsilon}_{E}+\widetilde{\mathbf{n}}\cdot\bar{\boldsymbol{\mu}}_{E}
	+\boldsymbol{\epsilon}_{B}\cdot\widetilde{\mathbf{n}}+\widetilde{\mathbf{n}}\cdot\bar{\boldsymbol{\mu}}_{B}\cdot\widetilde{\mathbf{n}} \right) \cdot \delta\mathbf{E}=0,
\end{equation}
where $\widetilde{n}_{ij}=-\epsilon_{ijk}n_k$ \footnote{The cross product of two vectors can be rewritten as a matrix-vector product. For $\mathbf{n}=(n_{1},n_{2},n_{3})$, we construct an associated matrix
$\tilde{\mathbf{n}}$ with $\tilde{n}_{ij}=-\epsilon_{ijk}n_{k}$. Then, the following identities hold:
$ \mathbf{n}\times\mathbf{A}=\tilde{\mathbf{n}}\cdot\mathbf{A}$, $(\mathbf{n}\times\mathbf{A})\mathbf{B}=(\tilde{\mathbf{n}}\cdot\mathbf{A})\mathbf{B}$, and	$(\mathbf{A}\mathbf{B})\cdot\tilde{\mathbf{n}}=-\mathbf{A}(\tilde{\mathbf{n}}\cdot\mathbf{B})$.}. Solving the equation consists of two steps:  solving $\mathrm{det}(\boldsymbol{\Lambda}(n))=0$ to find a value of $n$ and then solving $\boldsymbol{\Lambda}(n)\cdot \delta\mathbf{E}(n)=0$ to find $\delta\mathbf{E}(n)$ associated with the value. Usually multiple values of $n$ are obtained, implying multirefringence. The relation between the polarization vectors associated with different values of $n$ can hardly be  known before finding the vectors explicily because they refer to different $\boldsymbol{\Lambda}$ matrices. But the relation between the polarization vectors of a given value of $n$ may be known from the symmetry of $\boldsymbol{\Lambda}$.  To solve the equation, we need the explicit forms of  $\boldsymbol{\epsilon}_{E}$, $\boldsymbol{\epsilon}_{B}$, $\bar{\boldsymbol{\mu}}_{B}$, and $\bar{\boldsymbol{\mu}}_{E}$, which are derived for Plebanski-type Lagrangians in Section \ref{sec:pers}. From now on, $\delta\mathbf{E}$ refers to the probe electric field without the factor $\exp(i\omega(\mathbf{n}\cdot\mathbf{x}-t))$.

\section{Comparison with the formulation of magneto-electric effect in material science} \label{sec:me_material_science}

The 3+1 framework in Section \ref{sec:framework} closely follows the approach in conventional optics and thus can facilitate directly comparing the magneto-electric effects in field-modified vacua with those in optical media. Although the magneto-electric effects are hardly observed in usual optical media, some multi-ferroic materials have been known to show the effects. \cite{Fiebig2005Revival,Eerenstein2006Multiferroic,spaldin2019advances}  In this section, we compare the 3+1 framework with that used in material science, in which the material response is found by minimizing the Helmholtz free energy $\mathcal{F}$:\cite{Eerenstein2006Multiferroic}
 \begin{equation} \label{dPdM_material}
 \begin{split}
      \mathcal{F}(\mathbf{E},\mathbf{H})   = & - \frac{1}{2} \mathbf{E} \cdot \boldsymbol{\epsilon}\cdot \mathbf{E}  - \frac{1}{2} \mathbf{H} \cdot \boldsymbol{\mu} \cdot \mathbf{H} - \mathbf{E}\cdot \boldsymbol{\alpha} \cdot \mathbf{H} \\
     \mathbf{P}   = & - \frac{\partial \mathcal{F}}{\partial \mathbf{E}} = \boldsymbol{\epsilon} \cdot \mathbf{E} + \boldsymbol{\alpha} \cdot \mathbf{H} \\
      \mathbf{M} = & - \frac{\partial \mathcal{F}}{\partial \mathbf{H}} = \boldsymbol{\mu}\cdot \mathbf{H} +  \boldsymbol{\alpha}^T   \cdot \mathbf{E}, \\
 \end{split}
 \end{equation}
where $\boldsymbol{\epsilon}$ and $\boldsymbol{\mu}$ (symmetric) are permittivity and permeability tensors, respectively. The magneto-electric response tensor $\boldsymbol{\alpha}$ is odd under time reversal or space inversion to keep $\mathcal{F}(\mathbf{E},\mathbf{H})$ invariant under such transformations.\cite{Landau1984Electrodynamics} Here, all the quantities are real.

To compare (\ref{dPdM_material}) with (\ref{dDdH_dEdB}), we express $\delta \boldsymbol{\mathcal{P}}$ and $\delta \boldsymbol{\mathcal{M}}$ in (\ref{decomposition}) and (\ref{dDdH_dEdB}) in terms of $\delta\mathbf{E}$ and $\delta\mathbf{H}$. Substituting $\delta \mathbf{B} = \delta \mathbf{H} + \delta \boldsymbol{\mathcal{M}}$  in  (\ref{dDdH_dEdB}) while assuming that ${\bar{\boldsymbol{\mu}}}_{B}^{-1}$ exists, we obtain
\begin{equation} \label{dPdM_dEdH}
\begin{split}
	 \delta \boldsymbol{\mathcal{P}}  = & (\boldsymbol{\epsilon}_{E} - 1 -\boldsymbol{\epsilon}_{B} \cdot \bar{\boldsymbol{\mu}}_{B}^{-1} \cdot \bar{\boldsymbol{\mu}}_{E}) \cdot \delta \mathbf{E} + \boldsymbol{\epsilon}_{B} \cdot \bar{\boldsymbol{\mu}}_{B}^{-1} \cdot \delta \mathbf{H}, \\  
	 \delta \boldsymbol{\mathcal{M}}  = &(\bar{\boldsymbol{\mu}}_{B}^{-1}-1) \cdot \delta \mathbf{H} -  \bar{\boldsymbol{\mu}}_{B}^{-1} \cdot \bar{\boldsymbol{\mu}}_{E} \cdot \delta \mathbf{E}.
 \end{split}
\end{equation}
For the magneto-electric parts in (\ref{dPdM_dEdH})  to match those in  (\ref{dPdM_material}), $\boldsymbol{\epsilon}$, $\boldsymbol{\mu}$,  $\boldsymbol{\alpha}$ should be related to $\boldsymbol{\epsilon}_{E}$, $\boldsymbol{\epsilon}_{B}$, $\bar{\boldsymbol{\mu}}_{B}$, and $\bar{\boldsymbol{\mu}}_{E}$ as follows:
\begin{equation} \label{alpha_mu_epsilon}
\begin{split}
\boldsymbol{\epsilon} & = \boldsymbol{\epsilon}_{E} - 1 -\boldsymbol{\epsilon}_{B} \cdot \bar{\boldsymbol{\mu}}_{B}^{-1} \cdot \bar{\boldsymbol{\mu}}_{E},\quad \boldsymbol{\mu}  = \bar{\boldsymbol{\mu}}_{B}^{-1}-1 \\
\boldsymbol{\alpha} & = \boldsymbol{\epsilon}_{B} \cdot \bar{\boldsymbol{\mu}}_{B}^{-1}
	= - \left( \bar{\boldsymbol{\mu}}_{B}^{-1} \cdot \bar{\boldsymbol{\mu}}_{E} \right)^T. 
 \end{split}
\end{equation}
These relations and the symmetry of $\boldsymbol{\epsilon}$ and $\boldsymbol{\mu}$ lead to the relations 
\begin{equation} \label{epsE_muB_epsB}
\boldsymbol{\epsilon}_E = \boldsymbol{\epsilon}_E^T, \quad
\bar{\boldsymbol{\mu}}_B =  \bar{\boldsymbol{\mu}}_B^T, \quad 
\boldsymbol{\epsilon}_{B} = - \bar{\boldsymbol{\mu}}_{E}^T, 
\end{equation}
which also hold for the Plebanski-type Lagrangians, as will be shown in Section \ref{sec:3+1}.

The relation reciprocal to (\ref{alpha_mu_epsilon})  and (\ref{epsE_muB_epsB}) is obtained by expressing $\mathbf{P}$ and $\mathbf{M}$ in  (\ref{dPdM_material}) in terms of $\mathbf{E}$ and $\mathbf{B}$, which establishes the equivalence between the linear response coefficients in material science literature and those in our 3+1 formulation:
\begin{equation}
\begin{split}
    \boldsymbol{\epsilon}_E &= \boldsymbol{\epsilon} - \boldsymbol{\alpha} \cdot (1+\boldsymbol{\mu})^{-1} \cdot \boldsymbol{\alpha}^T +1 \\
    \boldsymbol{\epsilon}_B &= \boldsymbol{\alpha} \cdot (1+\boldsymbol{\mu})^{-1}  \\
    \bar{\boldsymbol{\mu}}_B &=  (1+\boldsymbol{\mu})^{-1}\\
    \bar{\boldsymbol{\mu}}_E &= - (1+\boldsymbol{\mu})^{-1} \cdot \boldsymbol{\alpha}^T.
\end{split}
\end{equation}
By substituting these expressions in (\ref{Lambda_eps_mu})  and rearranging terms, we obtain a factorized form of  $\boldsymbol{\Lambda}$: 
\begin{equation} \label{Lambda_material}
	\boldsymbol{\Lambda} =1+\boldsymbol{\epsilon} - \left( \widetilde{\mathbf{n}} + \boldsymbol{\alpha} \right) \cdot (1+\boldsymbol{\mu})^{-1} \cdot   \left( \widetilde{\mathbf{n}} + \boldsymbol{\alpha} \right)^T,
\end{equation}
from which the property of  $\boldsymbol{\Lambda}^T=\boldsymbol{\Lambda}$ is obvious. Because  $\boldsymbol{\Lambda}$ is real symmetric, the polarization vectors associated with a value of $n$ can be made orthonormal to one another.\cite{Horn2013Matrix}  The factorization looks implicative, but the specific forms of $\boldsymbol{\epsilon}$, $\boldsymbol{\mu}$, and  $\boldsymbol{\alpha}$ would be necessary for a further analysis.

\section{Light propagation modes in nonlinear electrodynamics of Plebanski-type Lagrangians}
\label{sec:3+1}
In this section, we derive the general expressions of the light propagation modes in nonlinear electrodynamics that are described by Plebanski-type Lagrangians. The general framework of mode analysis in Section \ref{sec:framework} is used to find the refractive indices and polarization vectors for a given propagation direction. 

\subsection{Permittivity, permeability, magneto-electric response tensors for Plebanski-type Lagrangians} \label{sec:pers}
When a strong electromagnetic field modifies the vacuum, the corresponding Lagrangian should be an analytic function of the Maxwell scalar $F$ and the Maxwell pseudo-scalar $G$ to satisfy the Lorentz and gauge invariance. Such Lagrangians are called Plebanski-type Lagrangians:\cite{plebanski1970lectures}
\begin{equation} \label{plebanski}
	\mathcal{L}=\mathcal{L}(F,G)=-F+\mathcal{L'},
\end{equation}
where  $\mathcal{L}'$ represents the non-Maxwell part. Typically,  $\mathcal{L}(F,G)$ is an even function of $G$ to preserve the parity.  

In general, the Lagrangian may have an imaginary part, which leads to absorption of the probe light.\cite{Heyl1997Birefringence} For example, the HES Lagrangian has an imaginary part either when only an electric field exists or when the electric field has a component parallel to the magnetic field.  It becomes significant as the purely electric field or the electric field component along the magnetic field approaches or surpasses the critical field strength $E_c= m_e^2 / e = 1.32\times 10^{16}\ \mathrm{W/cm^2}$.\cite{Karbstein2020Probing} As a consequence, electron-positron pairs are produced by strong electromagnetic fields, the so-called Sauter-Schwinger pair production or vacuum breakdown.\cite{Sauter1931Uber,Schwinger1951Gauge} In the current work, we do not take such an imaginary part into account to assume stable media. Practically, the imaginary term can be neglected when the electric field is weaker than $E_c/3$ and the magnetic field. \cite{Kim2023Vacuum}

Once the Lagrangian is given, the polarization ($\boldsymbol{\mathcal{P}}$) and the magnetization ($\boldsymbol{\mathcal{M}}$) are obtained by differentiating the Lagrangian with the electric and magnetic fields:\cite{Heisenberg1936Folgerungen,Berestetskii1982Quantum}
\begin{equation} \label{DPHM}
	\mathbf{D}=\mathbf{E}+\boldsymbol{\mathcal{P}}=\frac{\partial\mathcal{L}}{\partial\mathbf{E}},
	\quad
	\mathbf{H}=\mathbf{B}-\boldsymbol{\mathcal{M}}=-\frac{\partial\mathcal{L}}{\partial\mathbf{B}}.
\end{equation}
For Plebanski-type Lagrangians, we obtain
\begin{equation} \label{DH_EB}
	 \mathbf{D} = -\mathcal{L}_{F}\mathbf{E}-\mathcal{L}_{G}\mathbf{B},
	 \quad
	 \mathbf{H} = -\mathcal{L}_{F}\mathbf{B}+\mathcal{L}_{G}\mathbf{E},
\end{equation}
where the subscripts $F$ and $G$ denote the partial differentiation of $\mathcal{L}$ with respect to them: e.g., $\mathcal{L}_{FG} \equiv \partial^{2}\mathcal{L}/{(\partial F\partial G )}$. 

Then, we can obtain the permittivity, permeability, and magneto-electric response tensors by taking a variation of  $\mathbf{E}$ and $\mathbf{B}$ in (\ref{DH_EB}) and applying  the decomposition argument in Section \ref{sec:framework}: 
\begin{equation} \label{eps_mu}	
	\begin{split}
		\boldsymbol{\epsilon}_{E} =& -L_{F}\mathbf{I} \\ 
  & +\left[L_{FF}\mathbf{E}_{0}\mathbf{E}_{0}+L_{FG}\left(\mathbf{E}_{0}\mathbf{B}_{0}+\mathbf{B}_{0}\mathbf{E}_{0}\right)+L_{GG}\mathbf{B}_{0}\mathbf{B}_{0}\right], \\
		\boldsymbol{\epsilon}_{B} =& -L_{G}\mathbf{I} \\
  & +\left[-L_{FF}\mathbf{E}_{0}\mathbf{B}_{0}+L_{FG}\left(\mathbf{E}_{0}\mathbf{E}_{0}-\mathbf{B}_{0}\mathbf{B}_{0}\right)+L_{GG}\mathbf{B}_{0}\mathbf{E}_{0}\right], \\
		\bar{\boldsymbol{\mu}}_{B} =& -L_{F}\mathbf{I} \\
  & +\left[-L_{FF}\mathbf{B}_{0}\mathbf{B}_{0}+L_{FG}\left(\mathbf{B}_{0}\mathbf{E}_{0}+\mathbf{E}_{0}\mathbf{B}_{0}\right)-L_{GG}\mathbf{E}_{0}\mathbf{E}_{0}\right], \\
	\bar{\boldsymbol{\mu}}_{E} = & L_{G}\mathbf{I} \\
 & +\left[L_{FF}\mathbf{B}_{0}\mathbf{E}_{0}+L_{FG}\left(\mathbf{B}_{0}\mathbf{B}_{0}-\mathbf{E}_{0}\mathbf{E}_{0}\right)-L_{GG}\mathbf{E}_{0}\mathbf{B}_{0}\right],
	\end{split}
\end{equation}
where the products of two vectors denote dyadics. The symbols $L_F$, $L_G$, $L_{FF}$, $L_{FG}$, and $L_{GG}$ refer to $\mathcal{L}_F$, $\mathcal{L}_G$, $\mathcal{L}_{FF}$, $\mathcal{L}_{FG}$, and $\mathcal{L}_{GG}$ evaluated with the condition $(\mathbf{E},\mathbf{B})=(\mathbf{E}_0,\mathbf{B}_0)$. From now on, we omit the subscript 0 not to clutter the notation. With these expressions of linear response tensors, we can solve the master equation (\ref{Lambda_eps_mu}) to find the light modes, as will be shown in Sec.\ref{sec:modes}. A similar formulation was presented by Robertson:\cite{Robertson2019Optical} the modes were analyzed for weakly nonlinear Lagrangians and a fixed propagation direction, and the coupling to axions were considered. 

The linear response tensors in (\ref{eps_mu}) exhibit some symmetries. First, $\boldsymbol{\epsilon}_E = \boldsymbol{\epsilon}_E^T$,  $\bar{\boldsymbol{\mu}}_B =  \bar{\boldsymbol{\mu}}_B^T$, and $
\boldsymbol{\epsilon}_{B} = - \bar{\boldsymbol{\mu}}_{E}^T$,  i.e., (\ref{epsE_muB_epsB}). Second, $\boldsymbol{\epsilon}_{E} \leftrightarrow \bar{\boldsymbol{\mu}}_{B}$ and $\boldsymbol{\epsilon}_{B} \leftrightarrow -\bar{\boldsymbol{\mu}}_{E}$ under the duality transformation $(\mathbf{E},\mathbf{B}) \rightarrow (-i\mathbf{B},i\mathbf{E})$, which keeps $F$, $G$, and $\mathcal{L}$ the same. Finally,  $\boldsymbol{\epsilon}_{E}$ and  $\bar{\boldsymbol{\mu}}_{B}$ have even parity, while  $\boldsymbol{\epsilon}_{B}$ and  $\bar{\boldsymbol{\mu}}_{E}$ have odd parity: $\mathcal{L}$ should be an even function of $G$, i.e., $\mathcal{L}(F,-G)=\mathcal{L}(F,G)$, to respect the inversion symmetry.

\subsection{Light modes: refractive indices and polarization vectors for an arbitrary propagation direction} \label{sec:modes}
In this section, we systematically solve the master equation (\ref{Lambda_eps_mu}) with the tensors (\ref{eps_mu}) to find the light modes for an arbitrary propagation direction. The corresponding $\boldsymbol{\Lambda}$ in  (\ref{Lambda_eps_mu})  is now given as
\begin{equation} \label{Lambda_APQ} 
    \begin{split}
    \boldsymbol{\Lambda}= & -L_{F}\left[\left(1-n^{2}\right)\mathbf{I}+\mathbf{n}\mathbf{n}\right] \\
    & +L_{FF}\mathbf{P}\mathbf{P}+L_{FG}\left(\mathbf{P}\mathbf{Q}+\mathbf{Q}\mathbf{P}\right)+L_{GG}\mathbf{Q}\mathbf{Q},
    \end{split}
\end{equation}
where
\begin{equation} \label{APQ}
	\mathbf{P}=\mathbf{E}+\mathbf{n}\times\mathbf{B}, \quad \mathbf{Q}=\mathbf{B}-\mathbf{n}\times\mathbf{E}.
\end{equation}
From this dyadic expression of  $\boldsymbol{\Lambda}$, it is clear that $\boldsymbol{\Lambda}$ is real symmetric as in the case of material science, which is a feature of Plebanski-type Lagrangians. Through an attempt to find complete squares of $\mathbf{P}$ and $\mathbf{Q}$ in $\boldsymbol{\Lambda}$, we express $\boldsymbol{\Lambda}$ as a linear combination of the identity matrix and self-conjugate dyadics, i.e., dyadics of two identical vectors:\cite{Gibbs1913Vector}
\begin{equation} \label{Lambda_XYZU}
	\boldsymbol{\Lambda}=\mathbf{X}\mathbf{X}+\mathbf{Y}\mathbf{Y}+\mathbf{Z}\mathbf{Z}+\mathbf{U},
\end{equation}
where
\begin{equation} \label{XYZU}
	\mathbf{X}=a\mathbf{P}+b\mathbf{Q}, \quad \mathbf{Y}=c\mathbf{Q}, \quad \mathbf{Z}=d\mathbf{n}, \quad \mathbf{U}=e\mathbf{I}
\end{equation}
and
\begin{equation} \label{abcde}
	\begin{split}
	&	a^2  = L_{FF},\quad  ab  =L_{FG},\quad  b^2  = \frac{L_{FG}^2}{L_{FF}},  \\
	&	c^2  = \frac{L_{FF} L_{GG} -L_{FG}^2}{L_{FF}},\quad   d^2  = -L_F, \quad  e  = -L_{F}\left(1-n^{2}\right).
	\end{split}
\end{equation}

This form of $\boldsymbol{\Lambda}$  (\ref{Lambda_XYZU}) implies that the reciprocal vectors of  $\mathbf{X}$, $\mathbf{Y}$, and $\mathbf{Z}$  may be used as basis vectors to represent $\delta\mathbf{E}$, as in crystallography. However, it requires non-degeneracy, i.e., the condition that the volume $V :=\mathbf{X} \times \mathbf{Y} \cdot \mathbf{Z}$ is not zero. The degeneracy ($V = 0$) is caused by either  $acd=0$ or $\mathbf{P}\times\mathbf{Q}\cdot \hat{\mathbf{n}}=0$, as can be found from (\ref{APQ}) and (\ref{XYZU}). The first condition $acd=0$ is independent of $\hat{\mathbf{n}}$ but depends only on the functional form of $\mathcal{L}(F,G)$ and the background field. In fact, the condition reduces to $c=0$ because $a^2 = L_{FF}$ and $d^2=-L_F$ do not vanish for non-trivial Lagrangians. An example of having $c=0$ is the MM Lagrangian, as will be shown in Section (\ref{MM Lagrangian}). In contrast, the second condition $\mathbf{P}\times\mathbf{Q}\cdot \hat{\mathbf{n}}=0$ depends also on $\hat{\mathbf{n}}$. The second condition is met by a configuration satisfying  $\mathbf{B}_{t}=\hat{\bf{n}}\times \mathbf{E}_{t}$,\footnote{For a direction vector $\hat{\mathbf{n}}$ and an arbitrary vector $\mathbf{A}$, we use the following notations: $A_n := \mathbf{A}\cdot\hat{\mathbf{n}}$, $\mathbf{A}_c := \mathbf{A}\times\hat{\mathbf{n}}$, $\mathbf{A}_t := (\hat{\mathbf{n}}\times\mathbf{A})\times\hat{\mathbf{n}}$, $A^2:=\mathbf{A}\cdot\mathbf{A}$, and $A_t^2 := A^2 - A_n^2$.} which we call the free propagation configuration (FPC). In FPC, $n$ is unity, and $\mathbf{S} := \mathbf{E} \times \mathbf{B}$ heads along $\hat{\mathbf{n}}$, and any vector perpendicular to $\hat{\mathbf{n}}$ is a polarization vector. In other words, the light in FPC  propagates as if it were in a free vacuum. In the non-generate case, the FPC is the only configuration with $n=1$. In the degenerate case of $c=0$, however, another configuration has $n=1$, as shown in Section~\ref{sec:degenerate case}. A special case of FPC is the collinear configuration: $\mathbf{E}\parallel\mathbf{B} \parallel \hat{\mathbf{n}}$. Below we deal with the non-degenerate case of $c\neq0$ first and then the degenerate case of $c=0$, disregarding the FPC due to its triviality.

\subsubsection{Non-degenerate case ($c\neq 0$)}
Non-degeneracy ($V \neq 0$) allows to represent $\delta\mathbf{E}$ as a linear combination of the reciprocal vectors of  $\mathbf{X}$, $\mathbf{Y}$, and $\mathbf{Z}$:
\begin{equation} \label{dE_pqr}
	\delta\mathbf{E}=p \mathbf{X}' + q \mathbf{Y}' + r \mathbf{Z}',
\end{equation}
where
\begin{equation} \label{XpYpZp}
	\mathbf{X}'= \frac{\mathbf{Y}\times\mathbf{Z}}{V},\quad \mathbf{Y}'= \frac{\mathbf{Z}\times\mathbf{X}}{V},\quad \mathbf{Z}'= \frac{\mathbf{X}\times\mathbf{Y}}{V}.
\end{equation}
Then the linear equations for $p$, $q$, and $r$ are obtained by multiplying (inner product) each of $\mathbf{X}$, $\mathbf{Y}$, and $\mathbf{Z}$  from the left to the master equation $\boldsymbol{\Lambda}\cdot \delta\mathbf{E}=0$ with $\boldsymbol{\Lambda}$ in (\ref{Lambda_XYZU}):
\begin{equation} \label{pol_vec_eq}
	M p+ H q=0, \quad 	H p+N q=0, \quad r=-\frac{\mathbf{Z}\cdot\mathbf{X}}{d^{2}}p-\frac{\mathbf{Z}\cdot\mathbf{Y}}{d^{2}}q,
\end{equation}
where
\begin{equation} \label{HMN}
\begin{split}
 H= & \left[\mathbf{X}\cdot\mathbf{Y}-\frac{(\mathbf{Z}\cdot\mathbf{X})(\mathbf{Z}\cdot\mathbf{Y})}{d^{2}}\right], \
	M= \left[X^{2}+e-\frac{(\mathbf{Z}\cdot\mathbf{X})^{2}}{d^{2}}\right], \\
	N= & \left[Y^{2}+e-\frac{(\mathbf{Z}\cdot\mathbf{Y})^{2}}{d^{2}}\right].
 \end{split}
\end{equation}
It is assumed that $d^2=-L_F \neq 0$, which holds in general because any nonlinear Lagrangian should have the Maxwell term, i.e., $-F$, as the leading term.

\begin{table*}[t]
	\centering
	\begin{tabular}{|c|c|}
		\hline
		Non-degenerate case ($c\neq 0$) & $\delta\mathbf{E}=p \mathbf{X}' + q \mathbf{Y}' + r \mathbf{Z}'$ \\
		$n (\neq1)$ from $MN-H^2=0$ (\ref{FresnelEq_red}) & $\mathbf{X}',\ \mathbf{Y}',\ \mathbf{Z}'$ from (\ref{XpYpZp})	 \\
		\hline
		cases &   $(p,q,r)$ \\
		\hline
		$H\neq0$, $M\neq0$, $N\neq0$ &  $ \left( H, -M, -(\mathbf{Z}\cdot\mathbf{X})H/{d^{2}} + (\mathbf{Z}\cdot\mathbf{Y})M/{d^{2}}  \right)$ \\
		 &  $\sim (-N, H, (\mathbf{Z}\cdot\mathbf{X})N/{d^{2}} - (\mathbf{Z}\cdot\mathbf{Y})H/{d^{2}} )$ \\
		$H=0$, $M\neq0$, $N=0$ &  $ (0, -M,  (\mathbf{Z}\cdot\mathbf{Y})M/{d^{2}}) \sim
		(0, 1,  -(\mathbf{Z}\cdot\mathbf{Y})/{d^{2}})$ \\
		$H=0$, $M=0$, $N\neq0$ &  $ (-N, 0, (\mathbf{Z}\cdot\mathbf{X})N/{d^{2}}) \sim (1, 0, -(\mathbf{Z}\cdot\mathbf{X})/{d^{2}}) $ \\
		$H=0$, $M=0$, $N=0$ &  $ \left(p,q,-(\mathbf{Z}\cdot\mathbf{X})p/{d^{2}} - (\mathbf{Z}\cdot\mathbf{Y})q/{d^{2}} \right); $ \\
		 & $p$ and $q$ are arbitrary. \\
		 \hline
		 \hline
		  Degenerate case ($c= 0$) & $\delta\mathbf{E}=p\mathbf{X}'+r\mathbf{Z}'$ \\
		  $n (\neq1)$ from $\widetilde{M} \widetilde{N}-\widetilde{H}^2=0$ (\ref{FresnelEq_dg}) & $\mathbf{X}',\ \mathbf{Z}'$ from (\ref{XpZp}) \\
		 \hline
		 cases & $(p,r)$ \\
		 \hline
		 $\widetilde{H}\neq0$, $\widetilde{M}\neq0$, $\widetilde{N}\neq0$ &  $ (\widetilde{H}, -\widetilde{M}) \sim (-\widetilde{N}, \widetilde{H})$ \\
		 $\widetilde{H}=0$, $\widetilde{M}\neq0$, $\widetilde{N}=0$ &  $ (0, -\widetilde{M} ) \sim (0,1) $ \\
		 $\widetilde{H}=0$, $\widetilde{M}=0$, $\widetilde{N}\neq0$ &  $ (-\widetilde{N}, 0) \sim (1,0) $ \\
		 $\widetilde{H}=0$, $\widetilde{M}=0$, $\widetilde{N}=0$ &   $(p,r)$; $p$ and $r$ are arbitrary. \\
		 \hline		
	\end{tabular}
	\caption{\label{tab:det pol vec} Determination of polarization vectors for a given refractive index. The symbol $\sim$ put between two quantities means that the two vectors are equivalent, i.e., differ by a factor. In this table, the case of $n=1$ is excluded. In the non-degenerate case, $n=1$ leads to the FPC. In the degenerate case, $n=1$ can hold when $\delta \mathbf{E} \propto \mathbf{Z}\times\mathbf{X}$ for general configurations as well as the FPC.}
\end{table*}

The linear equations of $p$, $q$, and $r$ (\ref{pol_vec_eq}) are solved in two steps. In the first step, the null determinant condition,  $MN-H^2=0$, is solved for $n$. In the second step, each $n$ is substituted into  (\ref{pol_vec_eq}), which is then solved for  $(p,q,r)$, shown in Table~\ref{tab:det pol vec}. In the particular case of $H=M=N=0$, $(p,q)$ can have arbitrary values, and the resulting polarization vector spans a plane:
\begin{equation}
	\delta \mathbf{E}= p \left( \mathbf{X}' - \frac{\mathbf{X}\cdot\mathbf{Z}}{d^2} \mathbf{Z}' \right)
	+ q \left( \mathbf{Y}' - \frac{\mathbf{Y}\cdot\mathbf{Z}}{d^2} \mathbf{Z}' \right)
\end{equation}

In contrast to the photon in a free vacuum, a photon in NED of Plebanski-type Lagrangian acquires a longitudinal component when $r\neq 0$: $\delta\mathbf{E} \cdot \mathbf{n} = r/d$ from (\ref{XYZU}), (\ref{dE_pqr}), and (\ref{XpYpZp}). Such a longitudinal component appears in anisotropic dielectric media.\cite{Born1999Principles}

Now we solve  $MN-H^2=0$ for refractive indices. Factorizing $MN-H^2$ by using (\ref{APQ}), (\ref{XYZU}), (\ref{abcde}), and (\ref{HMN}), we obtain the Fresnel equation:
\begin{equation} \label{FresnelEq}
	\left( \frac{\widetilde{P^2}}{n^2-1}-\lambda^{(+)} \right)
	\left( \frac{\widetilde{P^2}}{n^2-1}-\lambda^{(-)} \right) =0,
\end{equation}
where
\begin{equation} \label{lambda}
\begin{split}
	\widetilde{P^2}:= & P^2-n^2 P_n^2=E^2-2n S_{n} + n^2 (B_{t}^2-E_{n}^2), \\
	\lambda^{(\pm)} = &  \frac{-\beta \pm\sqrt{\beta^2-4\gamma\alpha}}{2\gamma}.
 \end{split}
\end{equation}
The parameters $\alpha$, $\beta$, and $\gamma$ are determined solely by the background field, regardless of $\hat{\mathbf{n}}$, and thus so is $\lambda^{(i)}$:
\begin{equation} \label{G-abc}
	\begin{split}
		\alpha & =  L_F^2 + 2 L_F \left(G L_{FG}- F L_{GG} \right)- \left( L_{FF}L_{GG}-L_{FG}^2 \right) G^2 \\
			  & = d^4 -2d^2 \left[ abG-(b^2+c^2)F \right] -a^2 c^2 G^2, \\
		\beta & =  L_F \left( L_{FF}+L_{GG} \right) - 2F \left( L_{FF}L_{GG}-L_{FG}^2 \right)  	\\
				 & = -d^2 (a^2+b^2+c^2) - 2 a^2 c^2 F, \\
		\gamma & = L_{FF}L_{GG}-L_{FG}^2 = a^2 c^2.
	\end{split}
\end{equation}
The equation (\ref{FresnelEq}) can be rewritten as
\begin{equation} \label{FresnelEq_red}
	(B_{t}^2-E_{n}^2-\lambda^{(\pm)}) n^2 -2 S_{n} n + E_{n}^2 + \lambda^{(\pm)} = 0,
\end{equation}
from which, we obtain the refraction indices:
\begin{equation} \label{n_sol}
	n^{(i)}_{\pm} = \frac{-S_{n} \pm \sqrt{S_{n}^2 + (\lambda^{(i)} - B_{t}^2+E_{n}^2)(\lambda^{(i)}+ E^2) }}{\lambda^{(i)} - B_{t}^2+E_{n}^2},
\end{equation}
where $i=\pm$. Without a loss of generality, we take only positive values of $n$ and solve (\ref{pol_vec_eq}) to obtain the corresponding polarization vectors (see Table~\ref{tab:det pol vec}). It should be mentioned that we may have multirefringence, even beyond birefringence, because (\ref{FresnelEq}) is a quartic equation of $n$.\cite{Lorenci2013Multirefringence} For a purely magnetic background field, $(n^{(i)})^2$ has a simpler form of
\begin{equation} \label{n2_B}
	( n^{(i)} )^2 = \left| \frac{\lambda^{(i)}}{\lambda^{(i)}-B_t^2} \right|,
\end{equation}
implying birefringence.

So far, we have implicitly assumed that $n \neq 1$. When $n = 1$, the dyadic $\boldsymbol{\Lambda}$ in (\ref{Lambda_XYZU}) becomes $\boldsymbol{\Lambda} = \mathbf{X}\mathbf{X} + \mathbf{Y}\mathbf{Y} +\mathbf{Z}\mathbf{Z}$ because of  $e=0$. Then, the condition of $MN-H^2=0$ reduces to $\mathbf{P}_t=\mathbf{Q}_t=0$, which is satisfied only for the FPC. In such a case, the light sees no effect of NED.

\subsubsection{Degenerate case ($c=0$)} \label{sec:degenerate case}
The condition $c=0$ nullifies $\mathbf{Y}$ in (\ref{XYZU}) and thus simplifies $\boldsymbol{\Lambda}$ Unless $\mathbf{Z} \parallel \mathbf{X}$, which turns out to be the FPC, we can proceed similarly as in the non-degenerate case by introducing two-dimensional reciprocal vectors:
\begin{equation} \label{XpZp}
	\mathbf{X}'= \frac{Z^2 \mathbf{X} - (\mathbf{X}\cdot\mathbf{Z})\mathbf{Z}}{Z^2 X^2 - (\mathbf{X}\cdot\mathbf{Z})^2}, \quad
	\mathbf{Z}'= \frac{X^2 \mathbf{Z} - (\mathbf{Z}\cdot\mathbf{X})\mathbf{X}}{X^2 Z^2 - (\mathbf{Z}\cdot\mathbf{X})^2},
\end{equation}
where $\mathbf{X}'$ and $\mathbf{Z}'$ span the same plane spanned by $\mathbf{X}$ and $\mathbf{Z}$; $\delta \mathbf{E}$ has no component perpendicular to the plane. The counterpart to the equations (\ref{dE_pqr}), (\ref{pol_vec_eq}), and (\ref{HMN}) becomes as follows:
\begin{equation} \label{eqs_dg}
	\begin{split}
	& \delta\mathbf{E} = p \mathbf{X}' + r \mathbf{Z}', \\
	& \widetilde{M} p+ \widetilde{H} r=0,  \quad 	\widetilde{H} p+\widetilde{N} r=0, \\
    &		\widetilde{H} =\mathbf{X}\cdot\mathbf{Z},\quad \widetilde{N}=Z^2+e, \quad 	 \widetilde{M}  =X^2+e.
	\end{split}
\end{equation}
The null determinant condition $\widetilde{M}\widetilde{N}-\widetilde{H}^2=0$ is given as a quadratic equation of $n$:
\begin{equation} \label{FresnelEq_dg}
	\begin{split}
	&	\left[ a^2 (E_{n}^2 - B_{t}^2 ) + b^2 ( B_{n}^2 - E_{t}^2) -L_F  -2abG \right] n^2 \\
	 & + 2 S_{n} (a^2+b^2) n -a^2 E^2 + 2abG -b^2 B^2 +L_F =0.
	 \end{split}
\end{equation}
Once $n$ is available from this equation, the polarization vectors can be found from $(p,r)$ in Table~\ref{tab:det pol vec}.

In contrast to the non-degenerate case, the degenerate case always has a non-FPC with $n=1$. When $n=1$, $\boldsymbol{\Lambda} = \mathbf{X}\mathbf{X}   +\mathbf{Z}\mathbf{Z}$. When $\mathbf{Z} \nparallel \mathbf{X}$ (non-FPC), $\mathbf{Z}\times\mathbf{X}$ becomes the polarization vector because $\boldsymbol{\Lambda}\cdot \delta \mathbf{E}=0$ is satisfied by construction. The case of $\mathbf{Z} \parallel \mathbf{X}$ corresponds to the FPC. An advantage of our 3+1 formulation is a transparent analysis of the degenerate case compared to that of the covariant formulation.

\subsection{Procedure to find modes in non-degenerate nonlinear electrodynamics} \label{sec:procedure}

The procedure to find light modes in non-degenerate nonlinear electrodynamics is outlined in Table~\ref{tab:procedure}.  Here, we succinctly present the essential formulas used for the procedure in order. It is assumed that the formulas for $\{L_F, L_{FF}, L_{FG}, L_{GG}\}$, and thus $\{a^2, ab, b^2, c^2, d^2\}$ in (\ref{abcde}) are known regardless of the field configuration.

\begin{table*}[t]
	\centering
	\begin{tabular}{|c|l|c|}
		\hline
		step & task & equations \\
		\hline
		1 & Specify the Lagrangian. & (\ref{plebanski})\\
		2 & Calculate $a^2$, $ab$, $b^2$, $c^2$, and $d^2$ without specifying $\mathbf{E}$ and $\mathbf{B}$. & (\ref{abcde})\\
		3 & Specify $\mathbf{E}$, $\mathbf{B}$, and $\hat{\mathbf{n}}$ and rule out the FPC. & \\
		4 & Calculate $a^2$, $ab$, $b^2$, $c^2$, and $d^2$ for $\mathbf{E}$ and $\mathbf{B}$, and confirm $c\neq0$. & (\ref{abcde})\\
		5 & Calculate $\alpha$, $\beta$, and $\gamma$ in terms of $a^2$, $ab$, $b^2$, $c^2$, and $d^2$. & (\ref{G-abc})\\
		6 & Calculate $\lambda^{(\pm)}$ and $n>0$. & (\ref{lambda}), (\ref{n_sol})\\
		7 & Calculate $\mathbf{X}$, $\mathbf{Y}$, $\mathbf{Z}$, $V$, $\mathbf{X}'$, $\mathbf{Y}'$, and $\mathbf{Z}'$. & (\ref{XYZU}), (\ref{XpYpZp})\\
		8 & Calculate $H$, $M$, and $N$. & (\ref{HMN}) \\
		9 & Determine $(p,q,r)$ and calculate $\delta \mathbf{E}$. & Table~\ref{tab:det pol vec}, (\ref{dE_pqr} )\\
		\hline		
	\end{tabular}
	\caption{\label{tab:procedure} Procedure to find the light modes in a nonlinear non-degenerate vacuum. The steps from 7 to 9 should be implemented for each value of $n$ found in step 6. In Section \ref{sec:procedure}, we present the essential formulas used in the procedure in order. A similar procedure can be set up in the degenerate case.}
\end{table*}

Given a configuration specified by $\{\mathbf{E}, \mathbf{B}, \hat{\mathbf{n}} \}$ in the 3+1 formulation, we then specialize $\{a^2, ab, b^2, c^2, d^2\}$ to the configuration and calculate $\{\alpha, \beta, \gamma \}$ in (\ref{G-abc}) to obtain the refractive index $n$ from  (\ref{n_sol}). The refractive index may be numerically evaluated if the Lagrangian has a complicated functional form. With the refractive index, we calculate $\mathbf{P}$ and $\mathbf{Q}$ in (\ref{APQ}) and the following auxiliary parameters:
\begin{equation} \label{conf_paras}
\begin{split}
	&	\mathbf{P}_c = \mathbf{E}_c + n \mathbf{B}_t, \quad \mathbf{Q}_c = \mathbf{B}_c -n\mathbf{E}_t, \\
	&	 \mathbf{P}\times\mathbf{Q} = \mathbf{S} - n(E^2+B^2)\hat{\mathbf{n}} + n(E_n \mathbf{E} + B_n \mathbf{B}) + n^2 S_n \hat{\mathbf{n}},  \\
	&	 \widetilde{PQ} := \mathbf{P}\cdot\mathbf{Q} - n^2 P_n Q_n = -(1-n^2)G, \\
	&	 \widetilde{P^2} := P^2 - n^2 P_n^2 = E^2 -2n S_n + n^2 (B_t^2 - E_n^2), \\
	&	 \widetilde{Q^2} := Q^2 - n^2 Q_n^2 = B^2 -2n S_n + n^2 (E_t^2 - B_n^2),
  \end{split}
\end{equation}
where the subscripts $n$ or $t$ denote components parallel or transverse to ${\bf n}$, respectively. 

Then $acH$, $M$, and $N$ in (\ref{HMN}) are expressed as
\begin{equation} \label{acHMN}
\begin{split}
		acH = &  a^2 c^2 \widetilde{PQ} + ab c^2 \widetilde{Q^2}, \\
		M = & a^2 \widetilde{P^2} + 2ab\widetilde{PQ} +b^2 \widetilde{Q^2} + d^2(1-n^2), \\
		N = & c^2 \widetilde{Q^2} + d^2 (1-n^2).
  \end{split}
\end{equation}
Unless $acH=M=N=0$, we form the following two vectors proportional to the polarization vectors in Table~\ref{tab:det pol vec}:
\begin{equation} \label{dE1_dE2}
\begin{split}
	\delta\mathbf{E}_1   = &  a^2 M \mathbf{P}_c +(acH+ab M) \mathbf{Q}_c \\ 
 & -\frac{(a^2 P_n + ab Q_n) acH-a^2 c^2 M Q_n}{d^2}   \mathbf{P}\times\mathbf{Q}, \\
	 \delta\mathbf{E}_2  = &  -\frac{acH a^2}{c^2} \mathbf{P}_c
	 -\left( a^2 N + \frac{acH ab}{c^2} \right) \mathbf{Q}_c \\ 
	 & +\frac{a^2 \left[ (a^2 P_n + ab Q_n)N-acH Q_n \right]}{d^2} \mathbf{P}\times\mathbf{Q}.
  \end{split}
\end{equation}
We can choose any non-zero vector from $\delta\mathbf{E}_1$ and $\delta\mathbf{E}_2$ as the polarization vector for the refractive index $n$. When both are non-zero, they differ from each other only by a multiplicative factor. When $acH=M=N=0$, the polarization vector spans a plane:
\begin{equation} \label{dE_plane}
\begin{split}
		\delta\mathbf{E}= & \left[ d^2 \mathbf{Q}_c - (a^2 P_n + ab Q_n) \mathbf{P}\times\mathbf{Q} \right] p \\
						 & + \left( a^2 d^2 \mathbf{P}_c + ab d^2 \mathbf{Q}_c + a^2 c^2 Q_n \mathbf{P}\times\mathbf{Q} \right) q,
       \end{split}
\end{equation}
where $p$ and $q$ are arbitrary real numbers.

In summary, light modes are completely determined by the formulas (\ref{G-abc}), (\ref{n_sol}), (\ref{acHMN}), (\ref{dE1_dE2}), and (\ref{dE_plane}) once $\{a^2, ab, b^2, c^2, d^2\}$ (\ref{abcde}) and the configuration parameters~(\ref{conf_paras}) are known.

\subsection{Comparison with covariant formulation} \label{sec:comp_cov}

The most economical way to obtain light modes is the covariant four-vector formulation, to which our 3+1 formulation is equivalent in this regard. With the photon propagation four-vector $k^\mu = \omega (1,\mathbf{n})$, the covariant vectors used for polarization tensors take the form
\begin{equation}
	F^{\mu\nu}k_{\nu}=\omega(\mathbf{E}\cdot\mathbf{n},\mathbf{P}), \quad
	F^{*\mu\nu}k_{\nu}=\omega(\mathbf{B}\cdot\mathbf{n},\mathbf{Q}).
\end{equation}
The Fresnel equation in the covariant formulation, for instance eq.~(12) of $f^2 := F^{~\alpha \mu} F_{\alpha}^{~~ \nu} k_{\mu} k_{\nu}$ in Lorenci et al.\cite{Lorenci2013Multirefringence} recovers eq.~(\ref{FresnelEq}) with the same $\{\alpha, \beta, \gamma \}$. 

However, the 3+1 formulation has a more direct connection to the usual concepts in optics, e.g., permittivity, permeability, and magneto-electric response tensors (\ref{eps_mu}), and provides the refractive indices and polarization vectors in terms of familiar quantities, $\mathbf{E}, \mathbf{B}$, and $\mathbf{\hat{n}}$. Thus, it can allow a richer comparison of the nonlinear vacua with conventional optical media. The formulation also reveals a unique mathematical structure of the nonlinear vacuum's response: the Fresnel matrix $\boldsymbol{\Lambda}$ consists of self-conjugate dyadics (\ref{Lambda_XYZU}), which facilitates the construction of polarization vectors. Lastly, the formulation is related to the 3+1 formulation in the curved spacetime,\cite{Komissarov2004Electrodynamics} which is popular in numerical simulation of general relativistic phenomena.\cite{Baumgarte2010Numerical} In our 3+1 formulation in the Minkowski spacetime, we use the Killing vector $\partial_t$ to decompose the spacetime into a one-parameter spacelike hypersurface $\Sigma_t$. Each observer on $\Sigma_t$ measures the energy-momentum of photons as $k^\mu$ and the electromagnetic field $\mathbf{E}, \mathbf{B}$.

\section{Application to various nonlinear electrodynamics Lagrangians}
\label{sec:models}
In this Section, we apply the 3+1 formulation to several NED Lagrangians for the case of purely magnetic background field: post-Maxwellian (PM),\cite{Ni2013Foundations} Born-Infeld (BI),\cite{Born1934Foundations} ModMax (MM),\cite{Bandos2020Nonlinear,Sorokin2022Introductory} and the Heisenberg-Euler-Schwinger (HES) QED\cite{Dittrich1976One} Lagrangians. The PM Lagrangians are frequently used in X-ray polarimetry for light propagation near pulsars and also in black hole physics. Recently, these NED Lagrangians and Plebanski-type Lagrangians, in general, have been actively used for theoretical black holes.\cite{Bronnikov2023Regular} Theoretical magnetic black holes whose magnetic field strength is at electro-weak scales have been proposed\cite{Maldacena2021Comments}, which requires QED corrections in the context of general relativity.  

\begin{figure}
	\centering
	\includegraphics[width=0.15\textwidth]{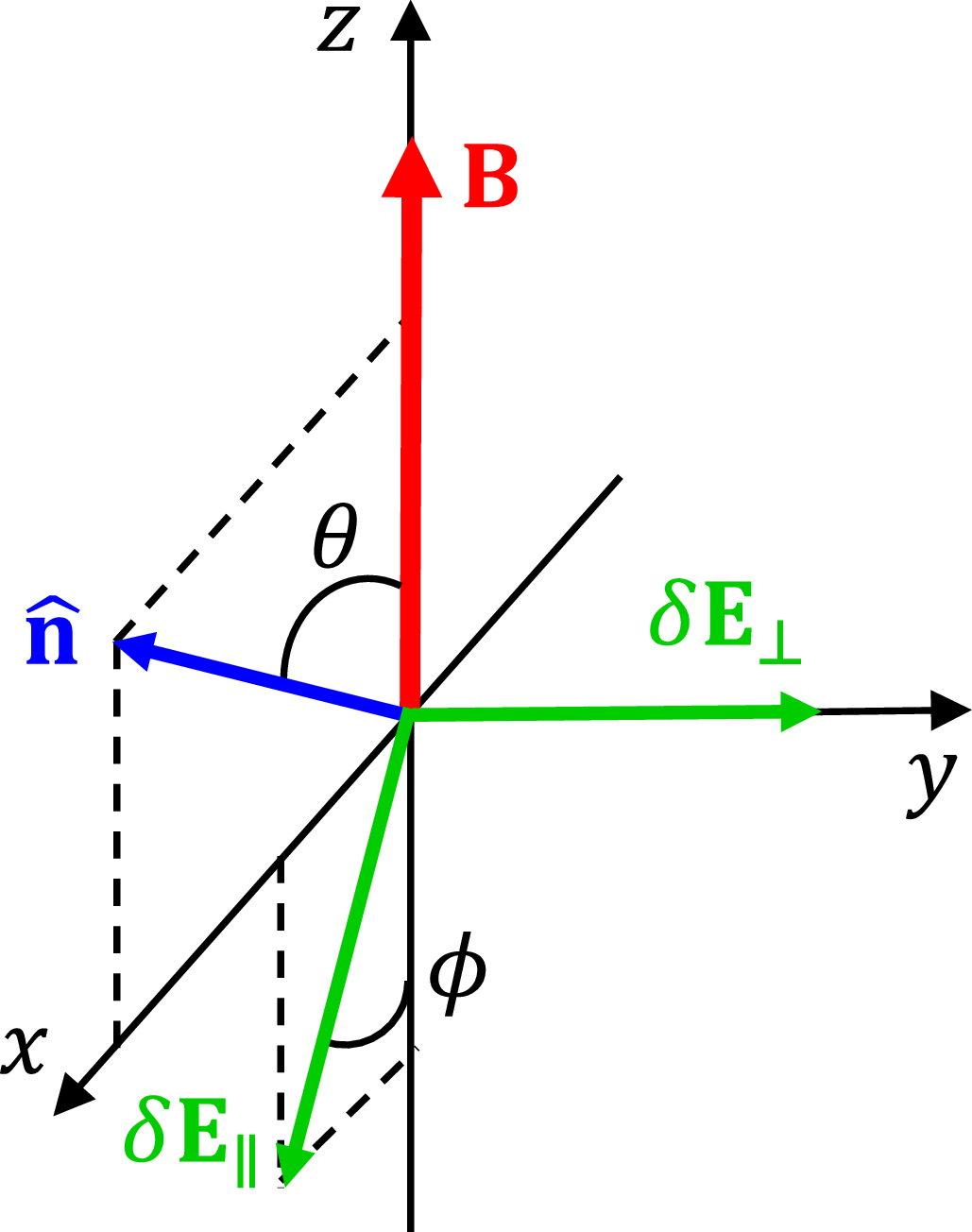}
	\caption{Configuration of the \textit{background magnetic field} ($\mathbf{B}=B_0 \mathbf{\hat{z}}$), the \textit{light's propagation direction} ($\mathbf{\hat{n}}$), and the two \textit{polarization vectors}  ($\delta\mathbf{E}_\perp$ and $\delta\mathbf{E}_\parallel$). $\delta\mathbf{E}_\perp$ is perpendicular to the plane formed by $\mathbf{B}$ and $\mathbf{\hat{n}}$, while $\delta\mathbf{E}_\parallel$ is on the plane.  The light modes in Table \ref{tab:modes_B} refers to this configuration.}
	\label{fig:conf}
\end{figure}

Without loss of generality, we assume that the magnetic field is along the $z$-axis, and the light's propagation direction vector lies on the $xz$-plane: $\mathbf{B}=B_0 \hat{\mathbf{z}}$ and $\hat{\mathbf{n}}=(\sin \theta,0,\cos \theta)$, as shown in Figure \ref{fig:conf}. The light modes for a given Lagrangian, i.e., $n$ and $\delta\mathbf{E}$, are found by following the procedure in Table~\ref{tab:procedure} and Section~\ref{sec:procedure}, as summarized in Table \ref{tab:modes_B}. 

Interestingly, the light propagation modes of the four Lagrangians exhibit a common structure, as manifestatively parametrized in Table \ref{tab:modes_B}, due to the fact that the Lagrangians are Plebanski-type. The polarization vector is either perpendicular to the plane  ($\perp$) formed by the magnetic field and the propagation direction or on the plane ($\parallel$). As a consequence of the common structure, the photon propagation at special angles is universal. When  $\theta=0$, $n_\perp=n_\parallel=1$  with   $\delta\mathbf{E}_\parallel=(1,0,0)$: the photon propagating along the magnetic field sees no background field effect. Note that $\delta\mathbf{E}_\perp=(0,1,0)$ regardless of $\theta$. When $\theta=\pi/2$, $\delta\mathbf{E}_\parallel=(0,0,1)$: the photon propagating perpendicular to the magnetic field has no longitudinal component. In general, non-zero $\epsilon$ implies the presence of a longitudinal component: $\delta\mathbf{E}_\parallel \cdot \mathbf{\hat{n}}=(\sin \theta \cos \theta) \epsilon / (1+\epsilon)$. The details specific to each Lagrangian are discussed below.

\subsection{Post-Maxwellian Lagrangian} \label{PM Lagrangian}
The PM Lagrangian \cite{Ni2013Foundations} is given as
\begin{equation} \label{LPM}
	\mathcal{L}_{\mathrm{PM}}=-F+ \eta_1 F^2 + \eta_2 G^2,
\end{equation}
where $\eta_1$ and $\eta_2$ are parameters. It is useful as the generic weak-field form of the Plebanski Lagrangians that approaches the Maxwell Lagrangian ($-F$) as the field vanishes. For example,  $\eta_1 = \eta_2 = 1/(2T)$ and  $\eta_1/4 =\eta_2/7 =e^4 /(360\pi^2 m^4)$  gives the BI and HES\cite{Heisenberg1936Folgerungen,Weisskopf1936Ueber,Schwinger1951Gauge} Lagrangians in the weak-field limit, respectively. It has been widely used to design vacuum birefringence experiments since only subcritical fields are available in the laboratory, even with the most intense lasers.\cite{King2016Measuring, Karbstein2020Probing, Yu2023X} One may extend $\mathcal{L}_{\mathrm{PM}}$ by adding a product of $F$ and $G$: a parity-violating term $\eta_3 FG$  or a parity-conserving term $\eta_3 F \sqrt{G^2}$.\cite{Ni2013Foundations}

The procedure in Table~\ref{tab:procedure} yields the coefficients (\ref{abcde}) and (\ref{G-abc}), regardless of the specific configuration:
\begin{equation}
	\begin{split}
	&	a^2 = 2 \eta_1, \ ab = 0, \     b^2 = 0, \ c^2 = 2 \eta_2, \   d^2 = 1-2 F\eta_1 \\
	&	\alpha = 4 F^2 \eta_1^2 -4(2F^2 +G^2)\eta_1 \eta_2 -4F (\eta_1 - \eta_2) +1  \\
	&	 \beta  = 4 F \eta_1^2 - 4F \eta_1 \eta_2 - 2(\eta_1 + \eta_2), \quad  \gamma = 4 \eta_1 \eta_2.
	 \end{split}
\end{equation}

The light modes under a pure magnetic field are given in Table~\ref{tab:modes_B}, which reduces in the weak-field limit to the well-known modes: $n_\perp  = 1 +  \eta_1 B_0^2   \sin^2 \theta $ with $\delta\mathbf{E}_\perp=(0,1,0)$  and $n_\parallel= 1 +   \eta_2 B_0^2   \sin^2 \theta $ with $\delta\mathbf{E}_\parallel=(\cos \theta,0,-\sin \theta)$.\cite{Adler1971Photon} The parallel mode has a longitudinal component: $\delta \mathbf{E}_\parallel \cdot \hat{\mathbf{n}} = \eta_2 B_0^2 \sin 2\theta $.

\begin{table*}[t]
	\centering
	\begin{tabular}{|c|c|c|c|c|}
		\hline
		Lagrangian  &degeneracy ($c=0$)&  $(\mathrm{refractive\ index})^2$& polarization vector&parameters\\
		\hline	
		\hline
		PM&No& $ n_\perp^2 \equiv \frac{1}{1-\mu \sin^{2}\theta} $& $\delta \mathbf{E}_\perp \equiv(0,1,0)$&$\mu = \frac{2\eta_1 B_0^2}{1-\eta_1 B_0^2}$\\
				  && $n_\parallel^2\equiv \frac{1+\epsilon}{1+\epsilon\cos^{2}\theta} $&
				 $\delta \mathbf{E}_\parallel \equiv \left(\cos\theta,0,- \frac{1}{1+\epsilon}\sin\theta\right) $&$\epsilon = \frac{2\eta_2 B_0^2}{1-\eta_1 B_0^2}$\\
		\hline
		 BI&No& $n_\parallel^2$& $p\delta \mathbf{E}_\perp + q \delta \mathbf{E}_\parallel$&$\epsilon=B_0^2/T$ \\
         & & & & $p$, $q$  arbitrary \\
		 \hline
		MM&Yes& $n_\perp^2(=1)$& $\delta \mathbf{E}_\perp $&$\mu=0$\\
		  && $n_\parallel^2$& $\delta \mathbf{E}_\parallel $&$\epsilon=e^{2g}-1$\\
		 \hline
 HES&No&  $n_\perp^2$&$\delta \mathbf{E}_\perp$&$\mu= B_0^2 a^2 /d^2$\\
  && $n_\parallel^2$&$\delta \mathbf{E}_\parallel$&$\epsilon=B_0^2 c^2 /d^2$\\
 \hline
	\end{tabular}
	\caption{ Refractive indices and polarization vectors for various Lagrangians in a magnetic background field: $\mathbf{B}=B_0 \hat{\mathbf{z}}$ and $\hat{\mathbf{n}}=(\sin \theta,0,\cos \theta)$, as shown in Fig.~\ref{fig:conf}. The acronyms of the Lagrangians are as follows: PM for post-Maxwellian, BI for Born-Infeld, MM for ModMax, and HES for  Heisenberg-Euler-Schwinger. For HES,  $a^2$, $c^2$, and $d^2$ are from  (\ref{HES:abcd}).   The FPC is not included here. }
	\label{tab:modes_B}
\end{table*}

\subsection{Born-Infeld Lagrangian} \label{BI Lagrangian}

The BI Lagrangian, introduced to remove the self-energy divergence in classical electrodynamics, is given as\cite{Born1934Foundations}
\begin{equation}
	\mathcal{L}_{\mathrm{BI}}=T-\sqrt{T^2+2 F T-G^2},
\end{equation}
where $T$ is a parameter. In the weak-field limit ($ T
\gg F, G$), the BI Lagrangian becomes the PM Lagrangian with $\eta_1 = \eta_2 = 1/(2T)$.

The coefficients (\ref{abcde}), (\ref{G-abc}), (\ref{lambda}) are given as
\begin{equation}
	\begin{split}
	& a^2 = \frac{T^2}{D^{3/2}},\quad ab = -\frac{GT}{D^{3/2}}, \quad     b^2 =
	\frac{G^2}{D^{3/2}}, \\
 & c^2 = \frac{1}{D^{1/2}}, \quad d^2 = \frac{T}{D^{1/2}} \\
&  \alpha = \frac{T^2 (2F+T)^2}{D^{2}},\quad   \beta=-\frac{2T^2 (2F+T)}{D^{2}},\quad
\gamma=\frac{T^2}{D^{2}}, \\
& \lambda^{(\pm)} = 2F+T
 \end{split}
\end{equation}
where  $D=T^2 +2FT-G^2$.

As shown in Table~\ref{tab:modes_B}, the BI Lagrangian does not have birefringence in a pure magnetic field.\cite{Sorokin2022Introductory} Furthermore, $H=M=N=0$ holds (see Table~\ref{tab:det pol vec}), and thus the polarization vector spans a plane as in a free vacuum but possibly with a non-zero longitudinal component: $\delta \mathbf{E} \cdot \hat{\mathbf{n}} =q\delta \mathbf{E}_\parallel \cdot \hat{\mathbf{n}} = q B_0^2 \sin 2\theta /(2T)$.

\subsection{ModMax Lagrangian} \label{MM Lagrangian}
The MM  Lagrangian, introduced as a modification of the Maxwell theory to preserve duality invariance and conformal invariance,\cite{Bandos2020Nonlinear} is given as
\begin{equation}
	\mathcal{L}_{\mathrm{MM}}=- F \cosh g + \sqrt{F^{2}+G^{2}} \sinh g,
\end{equation}
where $g\ge0$ for a lightlike or subluminal propagation of light. It reduces to the Maxwell Lagrangian when $g=0$. 

The coefficients (\ref{abcde}) are found as
\begin{equation}
	\begin{split}
	& a^2 = \frac{ G^{2}}{ R^3} \sinh g, \quad
	ab = - \frac{ FG}{ R^3} \sinh g, \quad
	b^2 = \frac{ F^{2}}{ R^3} \sinh g, \\	
	& c^2 = 0, \quad
	d^2 = \cosh g-\frac{ F}{R} \sinh g,
	\end{split}
\end{equation}
where $R=\sqrt{F^2+G^2}$. As $c^2=0$ (a consequence of the conformal and duality invariance), the MM Lagrangian is degenerate, and the formulas in Section \ref{sec:degenerate case} are used for analysis.

The light modes in a pure magnetic field are shown in  Table~\ref{tab:modes_B}. In contrast to the modes for other Lagrangians, the refractive indices and polarization vectors are independent of the strength of the background magnetic field, which is a consequence of conformal invariance.\cite{Sorokin2022Introductory} The perpendicular mode always has $n_\perp=1$ regardless of the configuration, as discussed in Section~\ref{sec:degenerate case}. A photon in the mode propagates as if it sees no background field.

\subsection{Heisenberg-Euler-Schwinger Lagrangian}
Heisenberg, Euler, and Schwinger obtained the one-loop correction to the Maxwell Lagrangian due to spin-1/2 fermions of mass $m$ in constant electric and magnetic fields of arbitrary strengths.\cite{Heisenberg1936Folgerungen,Schwinger1951Gauge} The exact one-loop action is given as the proper-time integral:
\begin{equation} \label{L1ab_int}
\begin{split}
& \mathcal{L}^{(1)}(h,g)  =-\frac{1}{8\pi^{2}}\int_{0}^{\infty}ds\frac{e^{-m^{2}s}}{s^{3}} \times  \\ 
 & \left\{ (ehs)\coth(ehs)(egs)\cot(egs) -\left[1+\frac{(ehs)^{2}-(egs)^{2}}{3}\right]\right\}  ,
\end{split}
\end{equation}
where
\begin{equation} \label{hg_FG}
h=\sqrt{\sqrt{F^{2}+G^{2}}+F},\quad g=\sqrt{\sqrt{F^{2}+G^{2}}-F}.
\end{equation}
The HES Lagrangian is obtained by adding the correction to the Maxwell Lagrangian:
\begin{equation}
\mathcal{L}_{\mathrm{HES}}(h,g)=\frac{g^{2}-h^{2}}{2}+\mathcal{L}^{(1)}(h,g).
\end{equation}

The direct analytical form of $\mathcal{L}_\mathrm{HES}$ for arbitrary constant electromagnetic fields is not known yet. However, the direct form for the case of $G=0$ was obtained by using the dimensional regularization,\cite{Dittrich1976One} and it was confirmed by the gamma-function regularization in the in-out formalism:\cite{Kim2019Quantum}
\begin{equation} \label{L_HES}
\begin{split}
	\mathcal{L}_\mathrm{HES}^{G=0}(\bar{h})  = &\frac{m^{4}}{8\pi^{2}\bar{h}^{2}}
	\left[ - \frac{\pi}{4\alpha_e} + \zeta'(-1,\bar{h}) \right] \\
 & + \frac{m^{4}}{8\pi^{2}\bar{h}^{2}} \left[ -\frac{1}{12}+\frac{\bar{h}^{2}}{4}-\left(\frac{1}{12}-\frac{\bar{h}}{2}
	+\frac{\bar{h}^{2}}{2}\right)\ln \bar{h}\right],
 \end{split}
\end{equation}
where $\bar{h}=m^2/(2eh)$, $\alpha_e=e^2/(4\pi)$ is the fine structure constant, $\zeta(s,\bar{h})$ is the Hurwitz zeta function, and $\zeta'(-1,\bar{h})=\mathrm{d}\zeta(s,\bar{h})/\mathrm{d}s\vert_{s=-1}$. For the case with $G\neq0$, the correction was obtained up to the order of $G^4$.\cite{Heyl1997Analytic} Recently, we expressed the correction as a power series in a small parameter ($g/h$).\cite{Kim2023Vacuum}

In finding the light modes in a pure magnetic field, one may be tempted to use the Lagrangian for a pure magnetic field, $\mathcal{L}_\mathrm{HES}^{G=0}(\bar{h}) \vert_{h=B_0}$. However, the Lagrangian is not sufficient, and the correction of $O(G)$ ( equivalently $O(g/h)$) should be included because the combined fields of the background field and the light field can have $G\neq0$ in general ($\delta\mathbf{E} \cdot \mathbf{B}_0$ is not necessarily zero). The correction is given as
\begin{equation}
	\mathcal{L}_\mathrm{cor}(\bar{h},\bar{g})= \frac{m^{4}}{8\pi^{2}\bar{h}^{2}} \frac{1}{\bar{g}^2}
	\left( -\frac{1}{24\bar{h}} + \frac{\ln \bar{h} - \psi^{(0)}}{12} \right),
\end{equation}
where $\bar{g}=m^2/(2eg)$, and $\psi^{(0)}(\bar{h})$ is the zeroth-order polygamma function.\cite{Kim2023Vacuum} Then the coefficients  $\{a^2, ab, b^2, c^2, d^2\}$  are obtained from (\ref{abcde}) by  applying  the condition $h=B_0$ and $g=0$ (purely magnetic background field):
\begin{equation} \label{HES:abcd}
\begin{split}
		 a^2  = &  \frac{16\alpha_{e}^{2}\bar{h}^{2}}{m^{4}}\left[2\bar{h}^{2}(\psi^{(0)}(\bar{h})-1)+\bar{h}\left(1-\ln\frac{\bar{h}\Gamma(\bar{h})^{2}}{2\pi}\right)+\frac{1}{3}\right], \\   ab  = & b^2=0, \\
		 c^{2}=& -\frac{8\alpha_{e}^{2}\bar{h}^2}{3m^{4}}\left[6\bar{h}^{2}-6\bar{h}\ln\frac{2\pi\bar{h}}{\Gamma(\bar{h})^{2}} \right] \\ 
  & -\frac{8\alpha_{e}^{2}\bar{h}^2}{3m^{4}} \left[ 1+2\psi^{(0)}(\bar{h})-24\zeta'(-1,\bar{h})+\frac{1}{\bar{h}}\right], \\
		 d^{2}= & \frac{\alpha_{e}}{6\pi}\left[6\bar{h}^{2}-6\bar{h}\ln\frac{2\pi\bar{h}}{\Gamma(\bar{h})^{2}}-24\zeta'(-1,\bar{h})+2\ln\bar{h}+1\right] + 1.
  \end{split}
\end{equation}
With $\{a^2, ab, b^2, c^2, d^2\}$ and $\{\mathbf{B}_0,\hat{\mathbf{n}}\}$, we find the light modes in the QED vacuum in an arbitrarily strong magnetic field, as shown in Table \ref{tab:modes_B}, by following the procedure in Table~\ref{tab:procedure} and Section \ref{sec:procedure}.  The simple expressions of the modes in Table \ref{tab:modes_B} were obtained by Melrose\cite{Melrose2013Quantum} and Denisov et al. \cite{Denisov2017Nonperturbative} albeit the coefficients $\mu$ and $\epsilon$ were given as complicated integrals instead of explicit functions as in (\ref{HES:abcd}).

\begin{figure}
	\centering
	\includegraphics[width=0.5\textwidth]{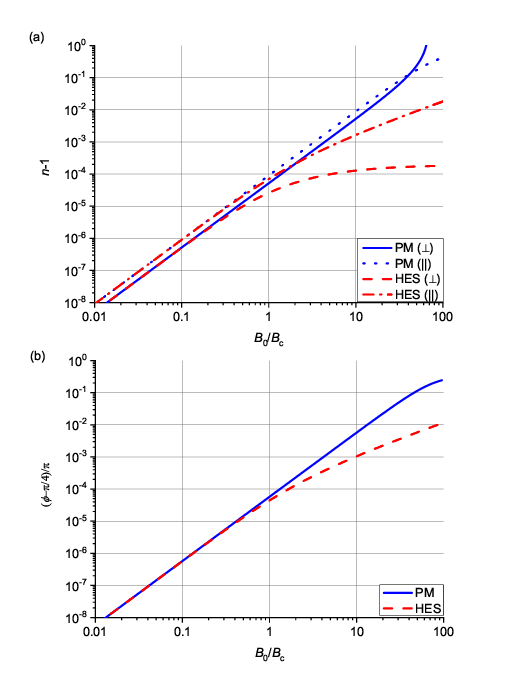}
	\caption{Variation of the \textit{refractive indices} ($n$) and the \textit{parallel polarization vector's angle} ($\phi$) from the free-space values as a function of the magnetic field strength for PM and HES Lagrangians: (a) $n_{\perp}-1$ and   $n_{\parallel}-1$, and (b) $(\phi-\pi/4)/\pi$. Here, the parameters $\eta_1$ and $\eta_2$ of PM are so chosen to match HES in the weak field limit:  $\eta_1/4 =\eta_2/7 =e^4 /(360\pi^2 m^4)$. $B_c = m^2 c^2/e \hbar = 4.4 \times 10^{13}\, {\rm G}$. The propagation vector's angle ($\theta$ in Fig.~\ref{fig:conf})  is $\pi/4$.}
	\label{fig:n_phi}
\end{figure}

As an illustrative example, we compare the light propagation modes of HES and PM Lagrangians under the assumption that HES approaches PM in the weak field limit.  In Figure \ref{fig:n_phi}, the departure of the refractive indices and the angle of the parallel polarization vector ($\phi$ in Figure \ref{fig:conf}) from their free-vacuum values are shown as a function of the magnetic field strength, and thus non-zero values represent the effect of the background magnetic field. As $B$ increases over about $0.5B_c$, the HES and PM modes begin to separate. At higher values of $B$, the background field effects for HES show a tendency of saturation compared to those for PM. The refractive index $n_\perp$ for PM diverges near $B=98.4B_c$ because the denominator in the expression of $n_\perp$ vanishes. These results imply that, in dealing with the light propagation around highly magnetized neutron stars (in particular, magnetars), the popularly used PM Lagrangian should be replaced with the HES Lagrangian because the magnetic field strength can surpass the critical field by two orders of magnitude in the vicinity of the magnetar's surface. Furthermore, such an environment accompanies a significant electric field component along the magnetic field. Such an electric field promotes the angle of the parallel polarization vector ($\phi$) to a sensitive measure of the background field effect in contrast to the electric field-free case.\cite{Kim2023Vacuum}  All these increasingly complicated situations can be analyzed by the  3+1 formulation in terms of the concepts familiar in conventional optics.

\section{Conclusion}
\label{sec:conclusion}

In this paper, we have provided a general formulation of the light modes in nonlinear electrodynamics, which are described by Plebanski-type Lagrangians, i.e., $\mathcal{L}=\mathcal{L}(F,G)$. Instead of Lorentz-covariant quantities, we used spatial vectors and dyadics (3+1 formulation) so that the formulation can be conveniently applicable to the situations of laboratory experiments or astrophysical observations. We found that nonlinear electrodynamics can be classified into two cases, non-degenerate and degenerate, in which the latter always possesses a mode with $n=1$. For each case, we derived general explicit formulas of refractive indices and polarization vectors for a given light propagation direction. For illustrations and practical applications, we applied the formulation to a magnetic background field described by the well-known NED Lagrangians, such as the post-Maxwellian, Born-Infeld, ModMax, and Heisenberg-Euler-Schwinger QED Lagrangians. We also advanced the possibility of comparing different Lagrangians in the weak magnetic field limit to test the underlying NED Lagrangians around extreme environments like highly magnetized neutron stars and black holes. Finally, we provided a streamlined procedure for determining light modes in practical applications (see Table \ref{tab:procedure} and  Section \ref{sec:procedure}).

By describing NED in the framework of conventional optics, the formulation proposed in this paper can be useful and more accessible for investigating light propagation in NED. The NED effects, i.e., the modification of the vacuum, may appear in various phenomena, of which the vacuum birefringence is the most promising to be observed because it is based on high-precision optical technologies. The terrestrial experiments employ strong permanent magnets \cite{Ejlli2020PVLASa} or ultra-intense lasers/x-ray free electron lasers \cite{Karbstein2021Vacuum,Yu2023X} to demonstrate vacuum birefringence. Although the strength of the background fields can be far below the critical field in the current or near future experiments,  the configuration can be designed as simple and convenient as possible for optimal observations.

In contrast, ongoing or proposed astrophysical observations of the x-rays from highly magnetized neutron stars, such as the Imaging X-ray Polarimetry Explorer (IXPE),\cite{Taverna2022Polarized} the enhanced X-ray Timing and Polarimetry (eXTP),\cite{Santangelo2019Physics} and the Compton Telescope project,\cite{Wadiasingh2019Magnetars} will probe the regime of supercritical magnetic fields,\cite{Olausen2014McGILL,Caiazzo2022Probing} albeit the configuration of background electromagnetic field is complicated and thus should be accurately modeled. As the combined electric and magnetic fields, for instance, the Goldreich-Julian dipole model, vary over macroscopic scale lengths, our 3+1 formulation can provide the local values of the refractive indices and polarization vectors. Then, from this local information, we should find the propagation path of light by solving the transport equations of rays and polarization vectors incorporating birefringence. The presented formulation will provide an essential element and concept for realistic model calculations. However, a realistic model for highly magnetized neutron stars should include plasma \cite{Wang2007Wave} and gravitational effects, which go beyond the scope of the present work and will be addressed in the future.

\begin{acknowledgments}
This work was supported by the Ultrashort Quantum Beam Facility operation program (140011) through APRI, GIST, and also by the Institute of Basic Science (IBS-R038-D1).  The authors are grateful to the anonymous reviewers for their constructive comments, which significantly improved the revision.
\end{acknowledgments}

\section*{Author Declarations}
\subsection*{Conflict of interest}
The authors have no conflicts to disclose.

\section*{Data Availability Statement}
The data that support the findings of this study are available from the corresponding author upon reasonable request.

%\appendix

\bibliography{refs_VB_master}% Produces the bibliography via BibTeX.

\end{document}